\documentclass[twocolumn,prb,a4paper,superscriptaddress,showpacs,showkeys]{revtex4}%
\usepackage[english]{babel}%
\usepackage{amssymb}%
\usepackage{amsmath}%
\usepackage{graphicx}%
\usepackage{hyperref}%
\usepackage{bm}%
\usepackage{xspace}%
\usepackage{color}%
\newcommand{\lem}{LE-$\mu$SR\xspace}%
\newcommand{\mean}[1]{\langle #1 \rangle}%
\begin{document}%
\title{Observation of non-exponential magnetic penetration profiles in the
       Meissner state --- A manifestation of non-local effects in superconductors}%
\date{\today}%
%
% Author list
%-----------------------------------------------------------------------%
%
\author{A. Suter}%
\email{andreas.suter@psi.ch}%
\homepage{http://lmu.web.psi.ch/lem/index.html}
\affiliation{Laboratory for Muon Spin Spectroscopy, Paul Scherrer Institute,
             CH-5232Villigen PSI, Switzerland}%
\author{E. Morenzoni}%
\email{elvezio.morenzoni@psi.ch}%
\affiliation{Laboratory for Muon Spin Spectroscopy, Paul Scherrer Institute,
             CH-5232Villigen PSI, Switzerland}%
\author{N. Garifianov}%
\affiliation{Laboratory for Muon Spin Spectroscopy, Paul Scherrer Institute,
             CH-5232Villigen PSI, Switzerland}%
\affiliation{Kazan Physical-Technical Institute, 420029 Kazan, Russian Federation}%
\author{R. Khasanov}%
\affiliation{Laboratory for Muon Spin Spectroscopy, Paul Scherrer Institute,
             CH-5232Villigen PSI, Switzerland}%
\affiliation{Physics Institute, University of Zurich, CH-8057 Zurich,
             Switzerland}%
\author{E. Kirk}%
\affiliation{Laboratory for Astrophysics, Paul Scherrer Institute, CH-5232
             Villigen PSI, Switzerland}%
\author{H. Luetkens}%
\affiliation{Laboratory for Muon Spin Spectroscopy, Paul Scherrer Institute,
             CH-5232Villigen PSI, Switzerland}%
\affiliation{Institut f\"{u}r Metallphysik und Nukleare
             Festk\"{o}rperphysik,  TU Braunschweig, 38106 Braunschweig, Germany}%
\author{T. Prokscha}%
\affiliation{Laboratory for Muon Spin Spectroscopy, Paul Scherrer Institute,
             CH-5232Villigen PSI, Switzerland}%
\author{M. Horisberger}%
\affiliation{Laboratory for Neutron Scattering, Paul Scherrer Institute, CH-5232
             Villigen PSI, Switzerland}%
%
% abstract
%----------------------------------------------------------------------------%
%
\begin{abstract}%
Implanting fully polarized low energy muons on the nanometer scale
beneath the surface of a superconductor in the Meissner state
enabled us to probe the evanescent magnetic field profile $B(z)$
($0<z\lesssim 200$nm measured from the surface). All the
investigated samples [Nb: $\kappa \simeq 0.7(2)$, Pb:$\kappa \simeq
0.6(1)$, Ta: $\kappa \simeq 0.5(2)$] show clear deviations from the
simple exponential $B(z)$ expected in the London limit, thus
revealing the non-local response of these superconductors. From a
quantitative analysis within the Pippard and BCS models the London
penetration depth $\lambda_{\rm L}$ is extracted. In the case of Pb
also the clean limit coherence length $\xi_0$ is obtained.
Furthermore we find that the temperature dependence of the magnetic
penetration depth follows closely the two-fluid expectation
$1/\lambda^2 \propto 1-(T/T_c)^4$. While $B(z)$ for Nb and Pb are
rather well described within the Pippard and BCS models, for Ta this is
only true to a lesser degree. We attribute this discrepancy to the
fact that the superfluid density is decreased by approaching the
surface on a length scale $\xi_0$. This effect, which is not taken
self-consistently into account in the mentioned models, should be
more pronounced in the lowest $\kappa$ regime consistently with our
findings.
\end{abstract}%
%
% pacs
%--------------------------------------------------------------------------------
%
% keywords
%----------------------------------------------------------------------------%
\maketitle%
%
% body
%--------------------------------------------------------------------------------
\section{Introduction}%
One of the fundamental properties of a superconductor is the
Meissner-Ochsenfeld effect\cite{meissner33}. It states that at low
magnetic fields and frequencies a superconductor expels or excludes
any magnetic flux from its core. However, at the surface the field
penetrates on a typical length scale $\lambda$ called the magnetic
penetration depth. In the London limit\cite{london35}, for a
semi-infinite superconductor, the functional $B$ vs.\ $z$ dependence
is an exponential one. This approximation holds for a large class of
superconductors. However, as found by Pippard from a set of
microwave experiments on impurity doped type-I
superconductors\cite{pippard53}, this is not always true. In analogy
with the anomalous skin effect Pippard introduced the concept of
non-local response of the superconductor, i.e. the screening current
trying to expel the magnetic field must be averaged over some
spatial region of the order of $\xi$ called the coherence length.
The physical interpretation of $\xi$ is, that it is the length over
which the superconducting wave function can be considered as rigid,
i.e.\ roughly speaking the size of a Cooper pair (for details see
Sec. \ref{sec:theory}). The non-local electrodynamical response
leads to various modifications of the London theory, one being that
the magnetic penetration profile $B(z)$ is no longer exponential and
even changes its sign beneath the surface of the superconductor. All
these findings were confirmed by the microscopic BCS
theory\cite{bcs57}.
Although these theoretical predictions have been known for half a
century, only very recently has a ``direct'' measurement of the
functional dependence of $B(z)$ been demonstrated\cite{suter04a}. A
historical summary of the experimental work on this subject begins
with that of Sommerhalder and
co-worker\cite{sommerhalder61a,sommerhalder61b,drangeid62} who
showed the existence of a sign reversal of $B(z)$ by measuring the
magnetic field leaking through a very thin superconducting tin film.
Doezema {\em et al.}\cite{doezema84} applied magnetoabsorption
resonance spectroscopy techniques to tackle the problem. The
technique uses the fact that quasi-particles traveling parallel to
the shielding current are bound to the surface by an effective
magnetic potential. Indication of non-local effects in Al were
inferred by comparing microwave induced resonant transitions between
the energy levels of these bound states with transition fields
calculated from the energy levels of the trapping potential,
parameterized to include the shape of the non-local BCS-like
potential. Due to the resonant character of the experiment, only a
few specific points of the potential are probed. In addition the
normal metallic state has to be understood very well in order to
interpret the data. This, together with uncertainties in modeling
the surface bound states, leaves room for speculations. Polarized
neutron reflectometry has also been applied since specular
reflectivity of neutrons spin polarized parallel or anti-parallel to
$\bm{B}$ depends on the field profile. However, this technique
requires model-fitting of spin-dependent scattering intensities
rather than giving a direct measure of the spatial variation of the
magnetic field. Up to now non-local corrections have been found to
lie beyond the sensitivity of polarized neutron reflectivity
techniques\cite{nutley94}.
In this paper we present magnetic field profiles measured beneath
the surface of various superconductors in the Meissner state by
means of low energy muon spin rotation spectroscopy (\lem). The
results provide a direct and quantifiable measure of non-local
effects in the investigated materials (Nb, Pb, Ta) and permit the
extraction of physical parameters such as the magnetic penetration depth
$\lambda$ and the coherence length $\xi$. The paper is organized as
follows: The next section reviews the theoretical framework
necessary to understand the results and the discussion. Section
\ref{sec:experimental} provides some information on the experimental
technique including the characterization of the samples. In
Sec.\xspace\ref{sec:results} we present our data, including a
discussion, followed by a summary in Sec.\xspace\ref{sec:summary}.

\section{Theoretical Background}\label{sec:theory}%

In this section, the theory of the response of a superconductor in
respect to an external electromagnetic field will be sketched (for a
rigorous derivation see Ref.\ \onlinecite{schrieffer64}). An
external electromagnetic field acts on the ground state of a
superconductor as a perturbation\footnote{This is at least true for
magnetic field strength $H \ll H_c$ and for frequencies $\nu \ll
2\Delta/\hbar$, were $\Delta$ is the energy gap and $\hbar$ the
Plank constant divided by $2\pi$.}. Within standard perturbation
expansion one can show\cite{schrieffer64} that the following
non-local relation between the current density $\bm{j}$ and the
vector potential $\nabla\wedge\bm{A} = \bm{B}$, where $\bm{B}$ is
the magnetic induction, holds:

\begin{equation}\label{eq:j_vs_A}
  j_\alpha(\bm{r}) = \sum_\beta \int \Big\{
  \underbrace{R_{\alpha\beta}(\bm{\rho})-
  \frac{e^2 n_{\rm S}}{m^*}\,\delta(\bm{\rho})\delta_{\alpha\beta}}_{\displaystyle  =:
  K_{\alpha\beta}(\bm{\rho})} \Big\}  A_\beta(\bm{r}')\, \text{d}\bm{r}'
\end{equation}

\noindent where $\bm{\rho} = \bm{r}-\bm{r}'$, $e$ is the charge,
$n_{\rm S}$ the supercurrent density, and $m^*$ the effective mass
of the charge carrier. $K_{\alpha\beta}(\bm{\rho})$ is called the
integral kernel. The vector potential $\bm{A}(\bm{r})$ needs to be
properly gauged in order that Eq.\ (\ref{eq:j_vs_A}) is physically
meaningful\footnote{Demanded gauge invariance is achievable by
requiring $\nabla\cdot\bm{A} = 0$, i.e.\ only the transverse part
of$\bm{A}$ has to be used.}. $R_{\alpha\beta}(\bm{\rho})$ describes
the paramagnetic response, whereas the second term in the bracket
reflects the diamagnetic one. If the ground state wave function of
the superconductor were ``rigid'' with respect to {\em all}
perturbations (rather than only those which lead to transverse
excitations) $R_{\alpha\beta}$ would be identically zero and Eq.\
(\ref{eq:j_vs_A}) would reduce to the second London equation

\begin{equation}\label{eq:london}
  j_\alpha(\bm{r}) = -\frac{1}{\mu_0}\,\frac{1}{\lambda_{\rm L}^2}\,
    A_\alpha(\bm{r})
\end{equation}

\noindent with the London penetration depth $\lambda_{\rm L} =
\sqrt{m^*/(\mu_0 e^2 n_{\rm S})}$, and $\mu_0$ the permeability of
the vacuum. This, together with the Maxwell equation
$\nabla\wedge\bm{B} = \mu_0 \bm{j}$, results, for a semi-infinite
sample, in the well known penetration profile

\begin{equation}\label{eq:Bz_exp}
  B(z) = B_{\text{ext}}\, \exp(-z/\lambda_{\rm L})
\end{equation}

\noindent where $z$ is the depth perpendicular to the surface and
$B_{\text{ext}}$ the externally applied magnetic field strength.

In situations where the paramagnetic term
$R_{\alpha\beta}(\bm{\rho})$ in Eq.\ (\ref{eq:j_vs_A}) cannot be
neglected one arrives at the more complicated formula

\begin{equation}\label{eq:Bz}
  B(z) = B_{\text{ext}}\, \frac{2}{\pi}\int \frac{q}{q^2+\mu_0 K(q\xi,T,\ell)}\,
    \sin(q z) \text{d}q.
\end{equation}

\noindent $K(q\xi,T,\ell)$ is the Fourier transformed kernel from
Eq.\ (\ref{eq:j_vs_A}) including the electron mean free path $\ell$.
Since only the one dimensional case will be considered everything is
expressed in scalar form. This equation reduces obviously to an
exponential decay if $K(q\xi,T,\ell)$ is independent of $q$, and in
the London limit $\mu_0 K(q\xi,T,\ell\to\infty) = 1/\lambda_{\rm
L}^2$. Eq.(\ref{eq:Bz}) is valid in the case of specular reflection
of the charge carriers at the surface. Another extreme limit is
given by assuming diffuse scattering at the interface. A real system
will not be exactly in one of the two limits. Since a quantitative
analysis\cite{tinkham80} shows that the difference in $\lambda$ is
marginal, we use Eq.(\ref{eq:Bz}) for the following discussion.
Still, we have verified by numerical integration of
Eq.(\ref{eq:j_vs_A}) that $B(z)$ for specular and diffuse scattering
are closely related.

Since the electromagnetic response of the superconductor is on the
length scale of $\lambda$, a local description would be sufficient
if the kernel $\mu_0 K(q\xi,T,\ell)$ is constant in the interval $0
\leq q \lesssim 1/\lambda$. In the next paragraphs we will show that
$\mu_0 K(q\xi,T,\ell)$ is approximately constant for $0\leq q
\lesssim 1/\xi$. From this it follows, that for $\lambda \gg \xi$
(type II) the local London limit should be quite reasonable, whereas
in the opposite limit $\lambda \ll \xi$ (type I) this is not the
case. For the type II case, non-local effects might play a role
either in case the energy gap exhibits nodes and henceforth $\xi$
will be very anisotropic\cite{kosztin97} or if $\kappa = \lambda /
\xi \lesssim 1.4$. We would like to stress that the border line
between the non-local and the local regime ($\kappa \approx 1.4$)
does {\em not} coincide with the border line between type-I and
type-II superconductivity ($\kappa = 1/\sqrt{2} = 0.7071\ldots$).

Knowing $K(q\xi,T,\ell)$ enables one to calculate $B(z)$. The
functional dependence of $K(q\xi,T,\ell)$ was derived
semi-phenomenologically by Pippard and later microscopically by
Bardeen, Cooper and Schrieffer.

\subsection{Pippard Kernel $K_{\rm
P}(q\xi,T,\ell)$}\label{sec:pippard}

Starting from a possible analogy between the anomalous skin effect
and the Meissner-Ochsenfeld effect, Pippard\cite{pippard53} arrived
at the formula for the kernel

\begin{figure}[h]
  \centering
  \includegraphics[width=\linewidth]{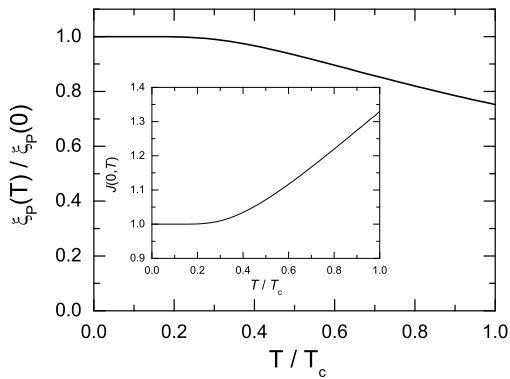}\\
  \caption{Temperature dependence of the Pippard coherence
           length $\xi_{\rm P}(T, \ell\to\infty)$ according to Eq.\
           (\ref{eq:xi_P}). The inset shows the temperature dependence
           of $J(0,T)$ within the weak coupling limit.}\label{fig:xi_P}
\end{figure}

\begin{widetext}
\begin{equation}\label{eq:KP}
  \mu_0 K_{\rm P}(q\xi, T,\ell) = \frac{1}{\lambda^2(T)}\,
  \frac{\xi_{\rm P}(T,\ell)}{\xi_{\rm P}(0,\ell)} \underbrace{\left[ \frac{3}{2}
  \frac{1}{x^3} \left\{\left(1+x^2 \right) \arctan(x) - x
  \right\}\right]}_{\displaystyle = g(x)},
\end{equation}
\end{widetext}

\noindent with $x = q \xi_{\rm P}(T, \ell)$, and the Pippard
coherence length $\xi_{\rm P}(T,\ell)$. The temperature dependence
of $\xi_{\rm P}(T, \ell)$ could be explained only later by the
BCS theory (see also the next section) and is

\begin{equation}\label{eq:xi_P}
  \frac{1}{\xi_{\rm P}(T,\ell)} = \frac{J(0,T)}{\xi_{\rm P}(0)} + \frac{1}{\ell}
\end{equation}

\noindent with

\begin{equation*}
   J(0,T) = \left( \frac{\lambda(T)}{\lambda(0)}\right)^2\,
                   \frac{\Delta(T)}{\Delta(0)}\,
                   \tanh\left[\frac{\Delta(T)}{2 k_{\rm
                   B}T}\right].
\end{equation*}

\noindent where $\Delta(T)$ is the superconducting energy
gap\cite{muehlschlegel59} and $k_{\rm B}$ the Boltzmann constant.
The weak temperature dependence of $\xi_{\rm P}(T,\ell\to\infty)$ is
shown in Fig.\ref{fig:xi_P}.

\subsection{BCS Kernel $K_{\rm BCS}(q\xi,T,\ell)$}\label{sec:bcs}

Starting from the weak coupling BCS model\cite{bcs57}, one arrives
at the following expression for the kernel\cite{halbritter71}

\begin{equation}\label{eq:K_bcs}
  \mu_0 K_{\rm BCS}(q\xi,T,\ell) = \sum_{n=0}^\infty \frac{1}{\Lambda_n(T,\ell)}\cdot g[q  \xi_n(T,\ell)]
\end{equation}

\noindent with the following set of abbreviations

\begin{eqnarray}\label{eq:Lambda_BCS}
  \Lambda_n(T,\ell) &=& \frac{1}{2a}\, \lambda_{\rm L}^2 f_n^3
      \left( 1 + \frac{\xi_n(T,\ell)}{\ell} \right) \nonumber \\
  \frac{1}{\xi_n(T,\ell)} &=&
      \frac{2}{\pi}\, \frac{f_n}{\xi_0}\,
      \frac{\Delta(T)}{\Delta(0)} + \frac{1}{\ell} \\
  a &=& \pi\, \frac{k_{\rm B}T}{\Delta(T)},
      ~~\xi_0 = \frac{\hbar v_{\rm F}}{\pi \Delta(0)} \nonumber \\
  f_n &=& \sqrt{1+(2n+1)^2 a^2} \nonumber \\
  \Delta(0) &=& \frac{\pi}{\gamma}\,
     k_{\rm B}T_c = 1.764\, k_{\rm B}T_c. \nonumber
\end{eqnarray}

\begin{figure}[h]
  \centering
  \includegraphics[width=0.9\linewidth]{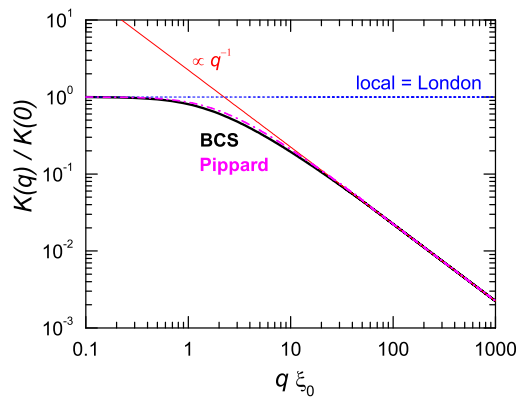}\\
  \caption{(Color online) $q$--dependence of the different kernels for $T\to 0$. The dashed
           line represents the $q$-independent London approximation.
           The BCS- (full) and Pippard-kernel (dash-dotted) are
           very similar and have a $q^{-1}$ asymptotic behavior
           for $q\to\infty$.}\label{fig:kernel}
\end{figure}

\noindent The temperature dependence of $\lambda$ is defined as
$1/\lambda^2(T) := \lim_{q\to 0} \mu_0 K(q\xi, T, \ell)$,
i.e.\xspace$\lambda(T\to 0,\ell\to\infty) = \lambda_{\rm L}$, and
$\xi_0$ the clean limit coherence length at $T=0$. Though the BCS
expression is much more involved, the $q$ dependence of the kernel
is {\em very} close to the one given in the phenomenological Pippard
expression Eq.\ (\ref{eq:KP}). A comparison is given in
Fig.\ref{fig:kernel}.

\begin{figure}
  \centering
  \includegraphics[width=\linewidth]{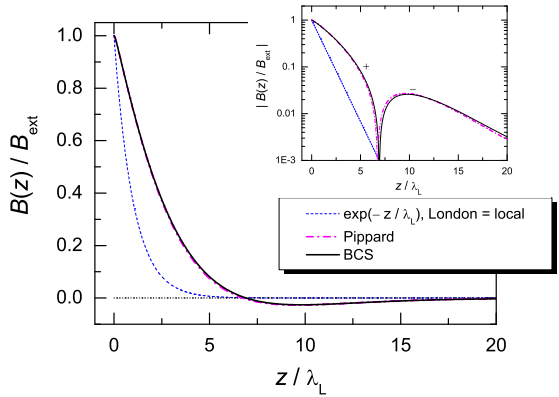}\\
  \caption{(Color online) The magnetic penetration profile in the Meissner state
           of Aluminum according to Eq.(\ref{eq:Bz}) for a
           $\xi_0 = 1600$~nm and $\lambda_{\rm L} = 50$~nm. The inset
           shows the same curve in a log scale.}\label{fig:hz_example}
\end{figure}

\noindent The corresponding magnetic fields can only be calculated
numerically. An example of an extreme type-I superconductor, Al, is
presented in Fig.\ref{fig:hz_example}. As intuitively expected
compared to the local case, the initial slope of $B(z)$ is reduced,
reflecting the fact that the magnetic field penetrates deeper into
the superconductor. The $\log\left| B(z)/B_{\text{ext}}\right|$
inset of Fig.\ref{fig:hz_example} shows initially a clear negative
curvature, i.e. a deviation from the exponential behavior. The next
important feature to be noticed is the sign reversal of the magnetic
field before approaching zero deep inside the sample. All these
findings can be made plausible by the following hand waving
arguments: In the non-local case the Cooper pairs are very extended
compared to the magnetic penetration profile. Since the partners
within a Cooper pair do not experience the same field, the screening
response is less effective and hence the slope is less steep
compared to a local response. This has a second effect: since the
field penetrates further beneath the surface, at some depth enough
Cooper pairs will experience it and start to ``overcompensate'',
which accounts for the negative curvature as well as for the field
reversal of $B(z)$ before approaching zero. This pictorial view is
clearly an oversimplification since the screening is due to the
Cooper pairs themselves, a strongly feedback coupled system, but it
provides the basis for a qualitative understanding of the
penetration profile.

\subsection{Strong Coupling Corrections}%

The BCS weak coupling theory assumes a weak electron-phonon
interaction. This is definitely not the case for either Pb or Nb.
Therefore, the much more elaborated strong-coupling
theory\cite{parks69} has to be used to calculate the properties of
the superconductor. A recent review describing the strong coupling
theory is found in Ref.\ \onlinecite{carbotte90}. Fortunately, the
outcome for the magnetic penetration profile $B(z)$ in the Meissner
state is almost the same. In particular Eq.(\ref{eq:Bz}) still holds,
and the structure of the kernel Eq.(\ref{eq:K_bcs}) is not altered,
except for a renormalization of $\xi_0$ and $\lambda_{\rm L}$ as
shown by Nam\cite{nam67}. These two length scales are renormalized
as

\begin{eqnarray}\label{eq:SC_renormalization}
  \lambda_{\rm L} &\to& \lambda_{\rm L} / \sqrt{Z} \\
  \xi_0           &\to& \xi_0 \cdot Z \nonumber
\end{eqnarray}

\noindent where the renormalization factors $Z$ for Pb, Nb and Ta
are $Z_{\rm Pb} \simeq 2.55$, $Z_{\rm Nb} \simeq 2.1$, and $Z_{\rm
Ta} \simeq 1.69$\cite{carbotte90}.

\section{Experimental Details}\label{sec:experimental}%

The samples investigated in this work were two Pb films, a Nb film,
and a Ta film. The characteristic properties, relevant for the
analysis are listed in Table \ref{tab:samples}.

\begin{table}[h]
  \centering
  \caption{Characteristics of the investigated samples.
                      $\mathrm{RRR} = [R(\mathrm{RT})-R(T_c+1.0)]/R(T_c+1.0)$
       }\label{tab:samples}
  \begin{ruledtabular}
  \begin{tabular}{c|c|c|c|c|c}
    sample & $B_{\rm ext}$ (G) & \multicolumn{2}{|c|}{thickness (nm)}
           & $T_c$ (K) & RRR \\
           &          & oxide layer & film & & \\\hline\hline
    Pb-I   & 88.2 & 16(2)  & 430(20)  & 7.1(1)  & 16 \\
    Pb-II  & 89.6 & 5.8(3) & 1055(50) & 7.21(1) & 23 \\
    Nb     & 88.2 & 4.2(3) & 310(15)  & 9.24(6) & 133 \\
    Ta     & 193.4 & 1.8(4) & 350(15)  & 4.42(2) & 45
  \end{tabular}
  \end{ruledtabular}
\end{table}

The samples were sputtered directly onto a sapphire crystal using
99.999 \% pure material. This permits both an excellent thermal
contact to the cold finger of the cryostat and also the application
of a bias to the sample, which is needed in the \lem experiments to
tune the implantation energy of the muons.
The Pb films were sputtered at room temperature, the base pressure
being $3\cdot 10^{-7}$~mbar. The Nb and Ta films were sputtered on
the substrate maintained at 1000~K. The base pressure in the
deposition chamber ranged between $2$ to $5\cdot 10^{-8}$ mbar.
X-ray diffractometry revealed epitaxial growth of Nb with a (2 0 0)
growth direction perpendicular to the substrate, and the Ta film
shows a highly oriented alpha structure most probably with a (2 0 0)
growth direction. The thickness of the films was determined by a
high sensitivity surface profiler and Rutherford backscattering. The
surface roughness of these films is given by a arithmetic mean
roughness value of $< 0.7$\%\footnote{The arithmetic mean rougness
value $R_a$ is defined as $R_a = 1/l_m \int_0^{l_m}
|y(x)|\mathrm{d}x$, where $y(x)$ is the variation perpendicular to
the averaged surface mean plane.}.
The critical temperature $T_{\rm c}$ was measured by means of
resistivity and susceptibility measurements. The mean free path
$\ell$ was estimated according to our resistivity data.

The oxide layers of Nb and Ta are extremely stable as discussed
extensively in Ref.\onlinecite{halbritter87}. The dieletric
$\mathrm{Nb_2O_5}$ forms at the surface of Nb and acts as a very
good protection layer. The typical $\mathrm{Nb_2O_5}$ layer
thickness found under the described growing conditions is $\simeq
5$~nm. Ta forms a $\mathrm{Ta_2O_5}$ oxide layer with a typical
saturation thickness of $\simeq 2$~nm, which again acts as a very
well protection layer for the Ta film. The thickness of the oxide
layer as determined by our measurements is in excellent agreement
with the findings in the literature.
\color{black}%__ASA__

The \lem method makes use of the muon spin rotation
technique\cite{schenck85} ($\mu$SR) where $\sim 100$\% polarized
positive muons ($\mu^+$) implanted in a solid sample rapidly
thermalize ($\sim 10$ ps) without noticeable polarization loss. The
spin evolution of the ensemble after the implantation is then
measured as a function of time. The evolution can be monitored by
using the fact that the parity violating muon decay is highly
anisotropic with the easily detectable positron emitted
preferentially in the direction of the $\mu^+$ spin at the moment of
the decay.

Counting the variation of the decay positron intensity $N(t)$ with
one or more detectors as a function of time after the muon has
stopped in the sample, it is possible to determine $P(t)$, the time
dependence of the polarization along the initial muon spin
direction.

\noindent The experimentally obtained time histograms have the form

\begin{equation}\label{eq:Nt}
  N(t) = N_0\, \left[ 1 + A_0\, \frac{P(t)}{P(0)} \right]\,
  e^{-t/\tau_\mu}.
\end{equation}

\noindent $N_0$ is a normalization constant reflecting the total
number of muons recorded. The exponential describes the decay of the
$\mu^+$ and $A_0$ is the maximum observable asymmetry of the decay
(theoretically $1/3$ in case of 100\% polarization and when
integrating over all positron energies). The relevant information
about the system under consideration is contained in the term $A_0\,
P(t)/P(0)$.

Unlike conventional $\mu$SR techniques which make use of
the energetic muons ($\sim 4$~MeV) originating from $\pi^+$ decay at
rest (``surface'' muons), \lem makes use of epithermal muons ($\sim
15$~eV) extracted after moderation of surface muons from a thin film
of a weakly bound van der Waals cryosolid (wide band gap
insulator)\cite{morenzoni94,morenzoni00}. By re-accelerating the
epithermal muons up to 20~keV and biasing the sample, it is possible
to tune the implantation energy in the range of 0.5 to 30 keV and
thus to implant the muons beneath the surface of any material in a
range of up to about 300~nm and a spatial resolution down to 1~nm.
Details concerning the $\mu^+$ stopping distribution will be given in
Sec.\ref{sec:results}.

The measurements were carried out as follows: A magnetic induction
$B_{\rm ext}$ parallel to the sample surface was applied after zero
field cooling the sample. The incident muon spin was parallel to the
surface and perpendicular to $B_{\rm ext}$.

\section{Results and Analysis}\label{sec:results}%

\begin{figure}[h]
  \centering
  \includegraphics[width=\linewidth]{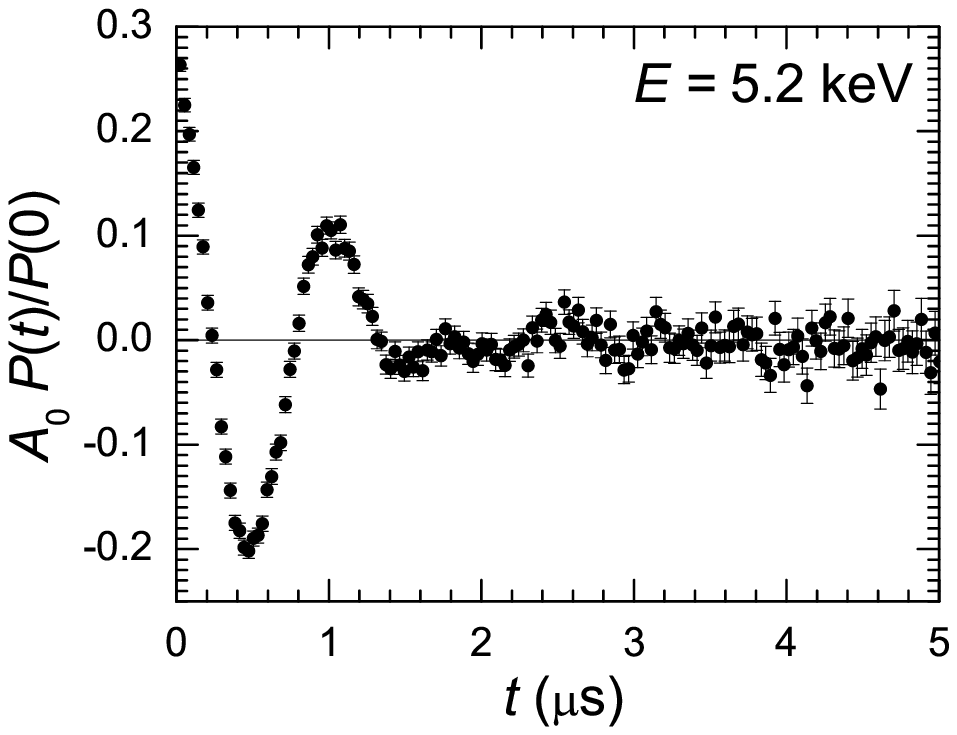}\\
  \includegraphics[width=\linewidth]{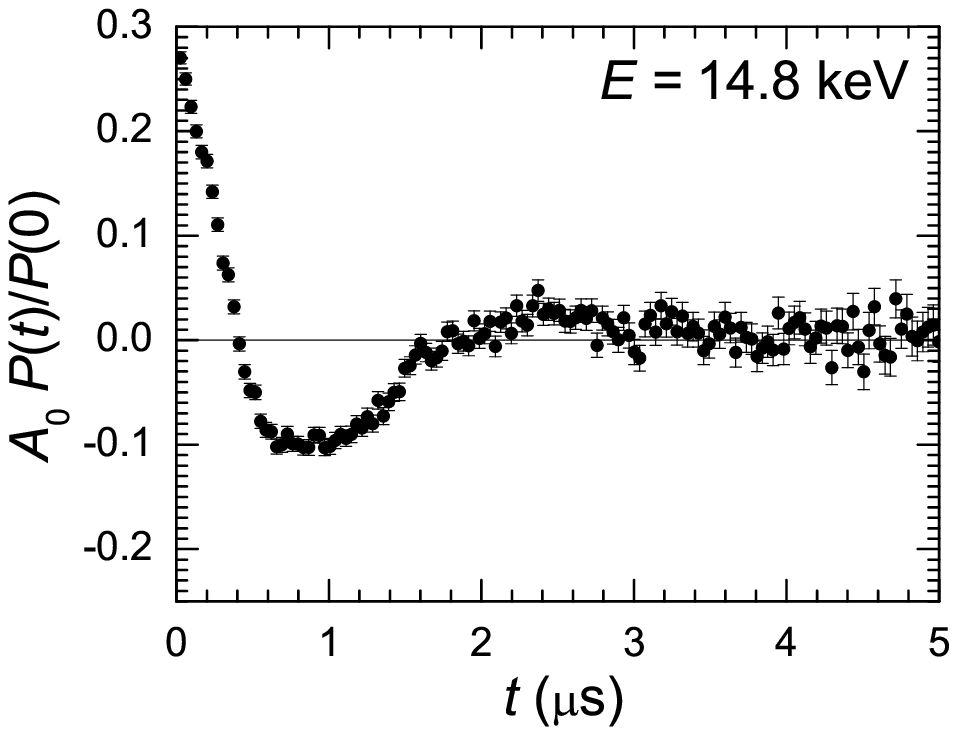}\\[-0.5cm]
  \caption{Typical time spectra $A_0\, P(t,E)/P(0,E)$.
           The shown data are from sample Pb-II, zero field cooled, $T=3.05$~K,
           $B_{\rm ext} = 8.82(6)$~mT. Top: $\mu^+$ implantation energy
           $E = 5.2$~keV. Bottom: $E = 14.8$~keV.}\label{fig:Pb_time_spectra}
\end{figure}

The applied magnetic field was determined from the Larmor precession
frequency of the muon spin at $T>T_c$. In this case the measured
polarization simply exhibits the undamped
precession\footnote{neglecting the contribution due to the small
nuclear damping} of the muon spin ensemble $P(t) \propto
\cos(\gamma_\mu B_{\rm ext} t + \phi)$. In the Meissner state
($T<T_c$), the $\mu^+$ ensemble, stopping in the surface layer, is
sampling $B(z)$, thus the muons will precess in various fields,
depending on where they actually stop (distance to the interface). This
yields a damped $P(t)$, reflecting the spatial distribution of
the magnetic field close to the surface.
Fig.\ref{fig:Pb_time_spectra} shows some typical time spectra
according to Eq.(\ref{eq:Nt}). From these spectra, the magnetic
field distribution $p(B,E)$ is obtained by Fourier transform

\begin{equation}\label{eq:Pt_pB}
  p(B,E) = \frac{2}{\sqrt{2\pi}} \int_0^\infty A_0\,
     \frac{P(t,E)}{P(0,E)}\, \cos(\gamma_\mu B t + \phi)\, \text{d}t
\end{equation}

\noindent where $\gamma_\mu$ is the gyromagnetic ratio of the
$\mu^+$ and $\phi$ an angle taking into account the concrete
detector geometry. $E$ is the implantation energy of the muons. To
obtain $p(B,E)$ we used a maximum entropy algorithm, which proved to
be much more robust than the usual Fourier transform methods,
especially in the case of limited available
statistics\cite{wu97,skilling84,rainford94,riseman00b}.

\begin{figure*}
  \centering
  \includegraphics[width=0.8\linewidth]{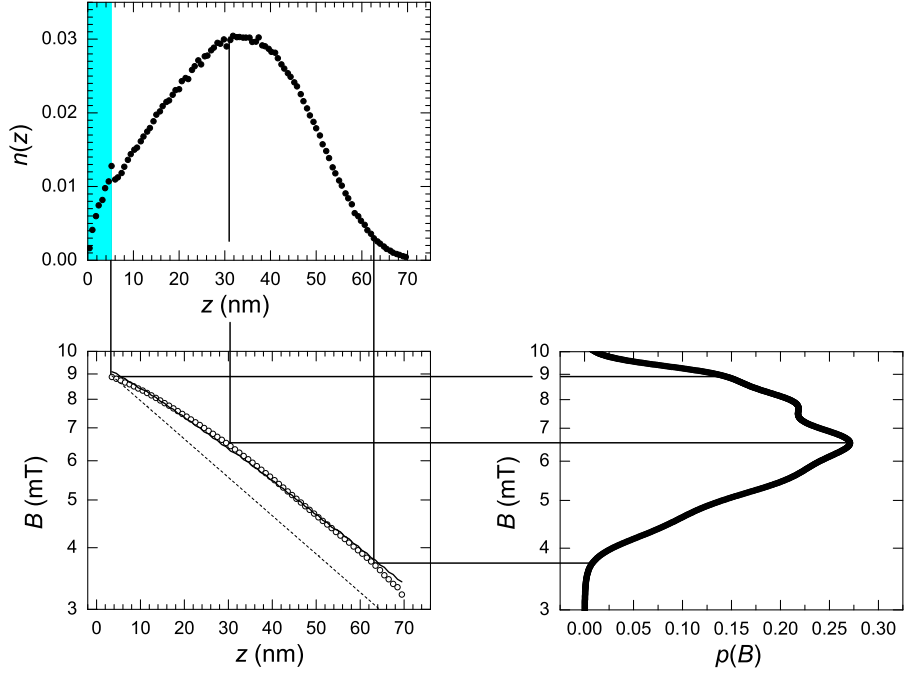}\\[-0.5cm]
  \caption{(Color online) Magnetic penetration profile $B(z)$ in sample
           Pb-II at $T=3.03$~K. The $\mu^+$ implantation energy
           was $E = 5.2$~keV. Top graph: $\mu^+$ stopping
           profile $n(z,E)$ from the Monte Carlo code {\tt TRIM.SP}.
           Bottom right graph: $p(B,E)$ maximum entropy analysis of
           $A_0\, P(t)/P(0)$ [see Eq.(\ref{eq:Nt})]. Bottom left graph:
           $B(z)$ field determination according to Eq.(\ref{eq:nz_pB_int}).
           The solid line in $B(z)$ is the best fit for the BCS kernel [Eqs.(\ref{eq:Bz}),
           (\ref{eq:K_bcs})], whereas the dashed line shows the London limit for
           parameters obtained from the BCS fit. The shaded area of $n(z,E)$
           shows the Pb oxide layer.}\label{fig:B_vs_z_reconstruction}
\end{figure*}

To determine $B(z)$ knowledge of the muon stopping distribution
$n(z,E)$ is needed. The situation is similar to the case of
magnetic resonance imaging where a measured nuclear resonance
frequency has to be translated into a space coordinate. We use the
Monte Carlo code {\tt TRIM.SP}\cite{eckstein91} whose reliability to
predict implantation profiles of low energy $\mu^+$ in various
materials has been confirmed by experimental testing\cite{morenzoni02}.
The functional relation between $n(z,E)$ and $p(B,E)$ is

\begin{equation}\label{eq:nz_pB}
  n(z,E)\, \text{d}z = p(B,E)\, \text{d}B
\end{equation}

\noindent which states nothing else than the probability that a given
implanted muon with incident energy $E$ stopping in the interval
$[z, z+\text{d}z]$, will experience a field in the interval $[B,
B+\text{d}B]$ with the probability $p(B,E)$ [assuming a monotonous
$B(z)$]. Integrating Eq.(\ref{eq:nz_pB}) on both sides yields

\begin{equation}\label{eq:nz_pB_int}
  \int_0^z n(\zeta, E)\, \text{d}\zeta = \int_{B(z)}^\infty
  p(\beta, E)\, \text{d}\beta
\end{equation}

\noindent which is, for a chosen $z$, an equation for $B$. Since
$n(z,E)$ can be calculated and $p(B,E)$ can be measured by means of
\lem, the magnetic field penetration profile can be determined.
Fig.\ref{fig:B_vs_z_reconstruction} shows an example of a $B(z)$
determination.

With this approach it is possible to determine the whole $B(z)$
functional dependence from a single measurement at one specific
implantation energy. Still, we determined $B(z)$ at various energies
which results in a set of overlapping curves. This self-consistence
check further demonstrates the reliability of the Monte Carlo code
used to determine $n(z,E)$. For a further crosscheck we used an
additional approach to determine $B(z)$, which allows a more
rigorous statistical error estimate. With a very narrow stopping
distribution, the following two approaches also lead to a $B(z)$:
(i) Plotting the spatial coordinate $z_p$ for which the stopping
distribution $n(z,E)$ is maximal (peak value), against the field
value $B_p$ where the field distribution $p(B,E)$ has its maximum, a
$B_p(z_p) = B(z)$ results for a set of different implantation
energies $E$. (ii) Instead of choosing the peak position, the mean
values $\mean{z} = \int z\, n(z,E)\, \text{d}z$ and $\mean{B} = \int
B\, p(B,E)\, \text{d}B$ can be used which again leads to $\mean{B}$
vs.\ $\mean{z} = B(z)$ for a set of different $E$. However, it has
to be taken into account that the stopping distribution $n(z,E)$ is
not very narrow and therefore a more elaborate calculation is
needed. The details of such an analysis are given in Appendix
\ref{app:B_vs_z_reconstruction}. The results can be summarized as
follows: Both approaches introduce minor systematic errors. The
systematic error of the peak value approach could mimic deviations
of an exponential decay reminiscent of non-local effects
$\{$negative initial curvature of $\log[B(z)]\}$ and therefore this
approach was excluded. The mean value determination is the
appropriate choice, since this method only produces systematic
errors {\em opposing} possible non-local effects, i.e.\ an
exponential decaying magnetic field profile is slightly deformed so
that $\log[B(z)]$ has a {\em positive} initial curvature. The
determination based on mean values could, in the worst case, only
lead to an underestimation of present non-local effects or even wash
them out.

The mean values $\mean{z}$ were determined directly from the Monte
Carlo stopping distribution. The asymmetries, measured at various
implantation energies, were fitted within a Gaussian relaxation
model to

\begin{eqnarray}\label{eq:At}
  A_0\, \frac{P(t)}{P(0)} &=&
          A_{\rm BG}^{\rm tot}\; \exp\left[-\frac{1}{2}\;(\sigma_{\rm BG} t)^2\right]
          \cos(\gamma_\mu B_{\rm ext} t + \phi) +  \nonumber \\
       && + A_{\rm SC}\; \exp\left[-\frac{1}{2}\;(\sigma_{\rm SC} t)^2\right]
          \cos(\gamma_\mu B_{\rm SC} t + \phi) - \nonumber \\
       && - A_{\rm BS}.
\end{eqnarray}

\begin{figure}
  \centering
  \includegraphics[width=0.9\linewidth]{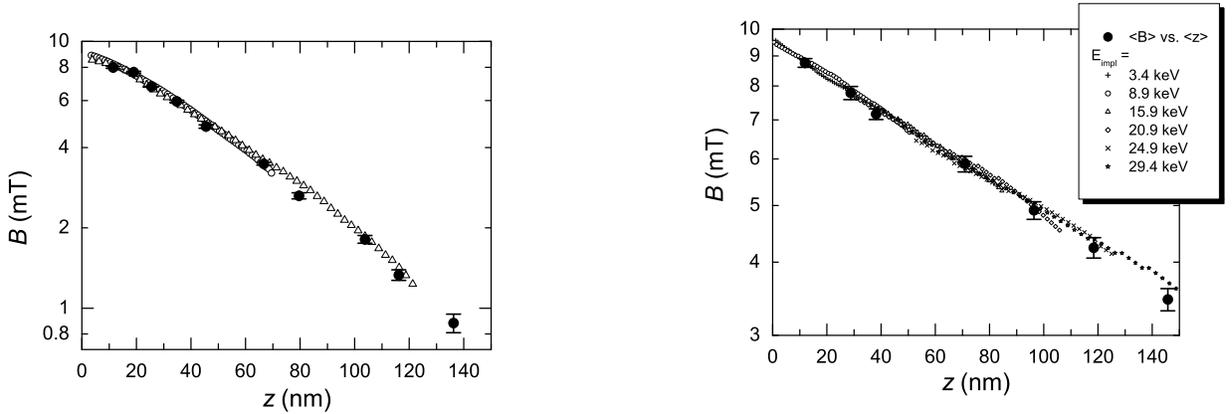}\\[-0.5cm]
  \caption{Mean value determination of $B(z)$ (solid dots) in sample Pb-II at
           $T=3.03$~K, together with integral approach [Eq.(\ref{eq:nz_pB_int})]
           for $E = 5.2$~keV (open circles) and $E = 14.8$~keV (open triangles).}\label{fig:B_vs_z_mean}
\end{figure}

\noindent $\phi$ is a phase describing the relative position of the
positron detectors. The observable asymmetry $A_0$ is modeled by 3
contributions $A_0=A_{\rm SC}+A_{\rm BG}^{\rm tot}+A_{\rm BS}$. For
our present experimental setup $A_0=0.27$. $A_{\rm SC}/A_0$ gives
the weight of the muons stopping in the superconductor. $A_{\rm
BG}^{\rm tot}/A_0$ is the portion of muons either stopping in the
oxide layer or in the sample surrounding. These muons experience the
external field $B_{\rm ext}$. $A_{\rm BS}/A_0$ is the portion of
backscattered muons. The backscattered muons form
muonium\cite{morenzoni02}  (a hydrogen like electron-muon bound
state with a gyromagnetic ratio $\gamma_{\rm Mu} \simeq 103
\gamma_{\mu}$). Due to its large $\gamma_{\rm Mu}$, muonium has a
much higher precession frequency compared to $\mu^+$ which results,
in the present experiment, in an instant depolarization (extremely
short dephasing time, filtering due to sampling of the time signal).
This leads to an effective reduction in the observable asymmetry as
written in Eq.(\ref{eq:At}). The various weights of the asymmetries
were fixed according to the Monte Carlo simulation of $n(z,E)$.
Therefore the only free fitting parameters are $B_{\rm SC}$, which
corresponds to $\mean{B}$ in this Gaussian relaxation model
approximation, and the two depolarization rates $\sigma_{\rm BG}$,
and $\sigma_{\rm SC}$. Fig.\ref{fig:B_vs_z_mean} shows a typical
result together with two $B(z)$ profiles determined from
Eq.(\ref{eq:nz_pB_int}) for different implantation energies. We
would like to stress that the mean value approach to determining
$B(z)$, due to slight systematic errors, rather underestimates the
deviation from an exponential magnetic penetration profile. Still,
deviations from the exponential behavior of $B(z)$ are clearly
present, thus confirming the non-local response.

\begin{figure}
  \centering
  \includegraphics[width=0.9\linewidth]{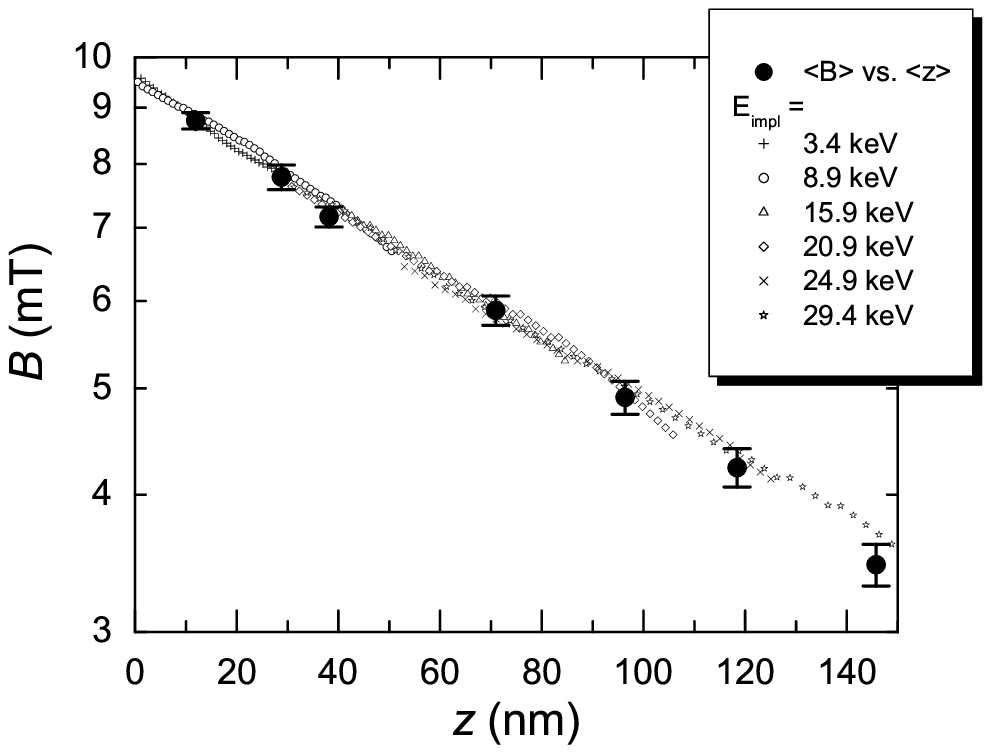}\\[-0.5cm]
  \caption{Magnetic penetration profile $B(z)$ of the high temperature
           superconductor $\mathrm{YBa_2Cu_3O_{7-\delta}}$ at $T=20$~K
           ($T_c = 87.5$~K). The $B(z)$ obtained from the mean values is
           displayed by solid circles whereas the other points show the integral
           determined field profile based on Eq.(\ref{eq:nz_pB_int}).}\label{fig:B_vs_z_YBCO}
\end{figure}

Additional evidence about the power to detect small variations of the
magnetic penetration profile is given by the previously measured
$B(z)$ data of the high temperature superconductor
$\mathrm{YBa_2Cu_3O_{7-\delta}}$ (optimally doped; for details see
Ref.\onlinecite{jackson00b}). In this clear cut type II
superconductor [$\xi_0\approx 1.5$~nm, $\lambda_{\rm L} =
146(3)$~nm] a perfect exponential $B(z)$ is expected, at least for
temperatures $\gtrsim 1$~K. Below $T \lesssim 1$~K there might be
deviations from the exponential behavior due to the d-wave pairing
character in this compound which can lead to substantial non-linear
and non-local effects\cite{yip92,kosztin97,amin98}. We reanalyzed
these data according to the above discussion. The results for
$T=20$~K are shown in Fig.\ref{fig:B_vs_z_YBCO}, which clearly
demonstrates the exponential form of $B(z)$.

\begin{figure*}
  \centering
  \includegraphics[width=0.45\linewidth]{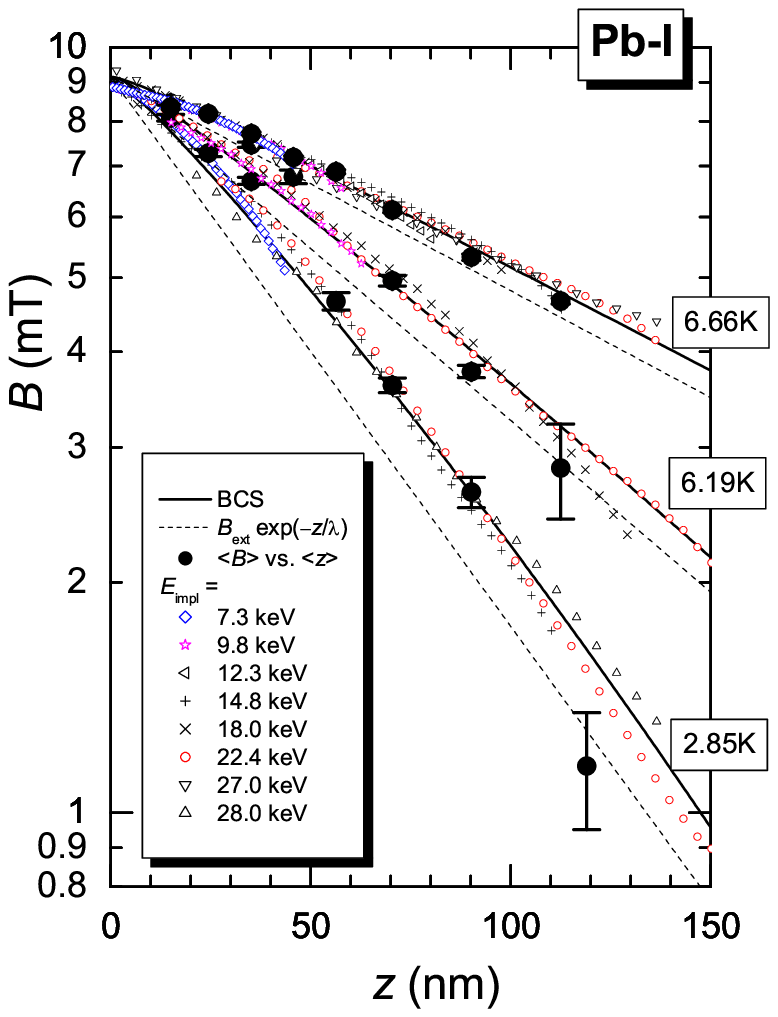} \qquad
  \includegraphics[width=0.45\linewidth]{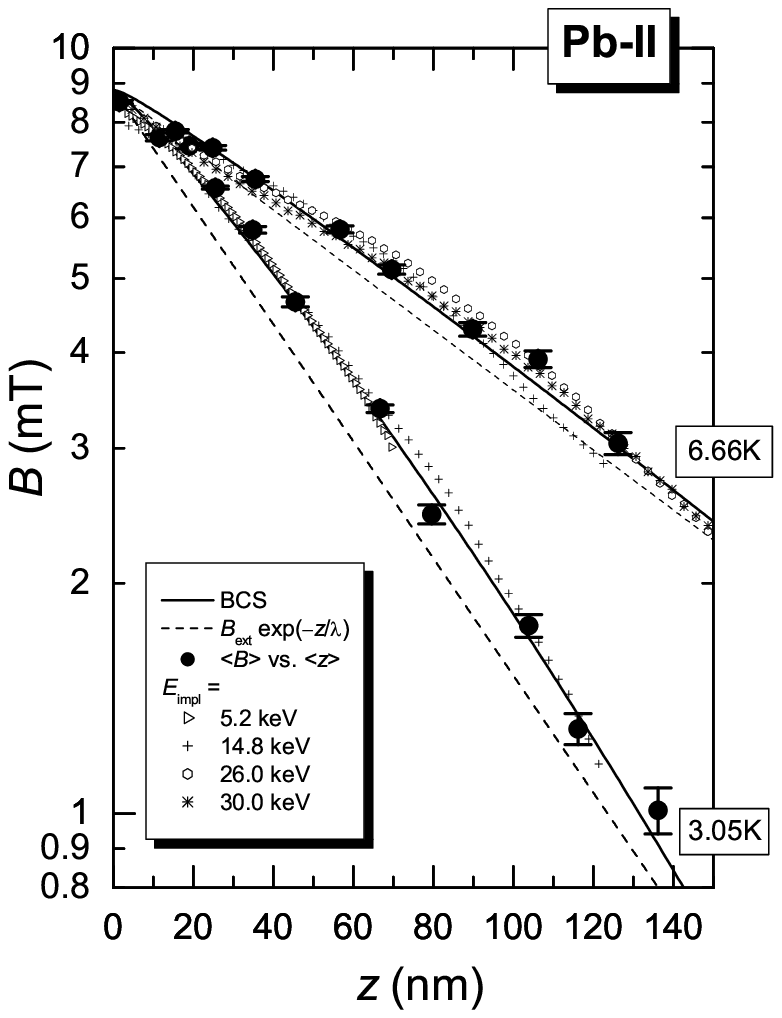}\\[-0.5cm]
  \caption{(Color online) Magnetic penetration profiles for Pb at various temperatures.
           Left graph: Data from sample Pb-I. Right graph: Data from sample Pb-II.
           The solid lines are BCS fits to the data, whereas the dashed line
           represents $B(z) = B_{\rm ext} \exp(- z/\lambda)$, where the
           $\lambda$ from the BCS fit is used.}\label{fig:Pb_B_vs_z}
\end{figure*}

\begin{figure}[h]
  \centering
  \includegraphics[width=\linewidth]{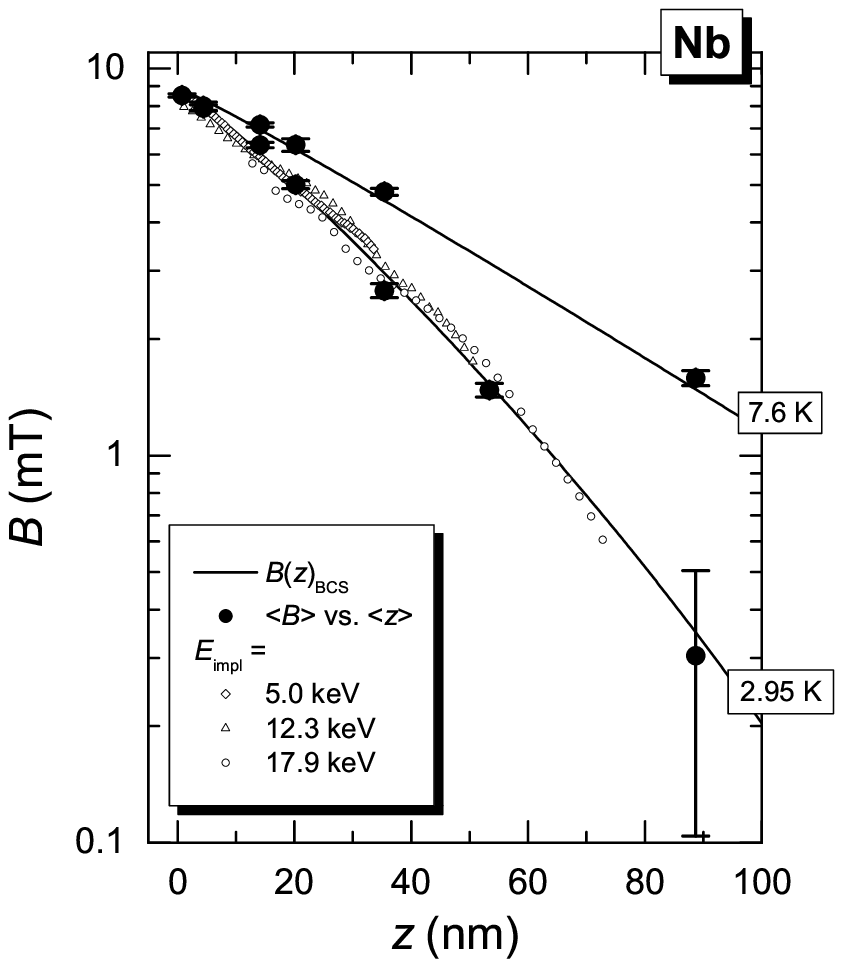}\\[-0.5cm]
  \caption{Mean value determination (solid dots) in Nb at
           $T=2.95$~K and $T=7.6$~K, together with the integral approach.
           The solid lines show BCS fits.}\label{fig:Nb_B_vs_z}
\end{figure}

\begin{figure}
  \centering
  \includegraphics[width=\linewidth]{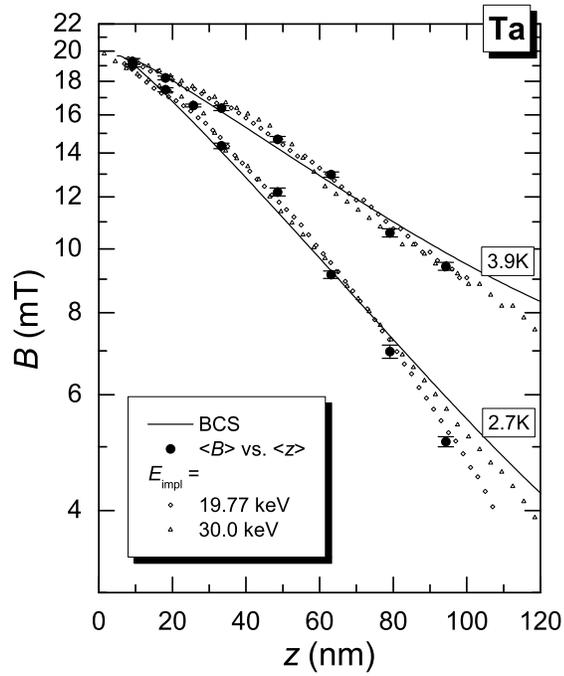}\\
  \caption{Magnetic penetration profiles for Ta at various temperatures.
           The solid lines show BCS fits.}\label{fig:Ta_B_vs_z}
\end{figure}

After having established the reliability of our approach to
determine $B(z)$, we turn to the discussion of the data.
Fig.\ref{fig:Pb_B_vs_z} shows a collection of our results for Pb;
Fig.\ref{fig:Nb_B_vs_z} shows the Nb and Fig.\ref{fig:Ta_B_vs_z} the
Ta data. The low temperature data of both Pb samples show a clear
deviation from an exponential decay law, indubitable evidence for
the presence of non-local effects. Furthermore, increasing the
temperature leads to less-pronounced curvature as expected,
since {\em very} close to $T_c$ non-local effects should disappear
altogether (strong temperature dependence of $\lambda$ for $T\to
T_c$; weak one of $\xi$ [see Fig.\ref{fig:xi_P}]). Unfortunately,
the range where a sign reversal of $B(z)$, predicted by theory,
should appear, is experimentally not accessible yet. To compare with
theory, the data were fitted according to Eq.(\ref{eq:Bz}) for the
Pippard kernel [Eq.(\ref{eq:KP})] and the BCS kernel
[Eqs.(\ref{eq:K_bcs}),(\ref{eq:Lambda_BCS})], respectively. Since
the temperature dependence of $\lambda$ is undefined within the
Pippard model, we chose the two-fluid approximation $\lambda(t) =
\lambda_0 / \sqrt{1-t^4}$, with $t=T/T_c$. $\lambda_0$ is an
effective London penetration depth, taking into account corrections
due to the scattering of electrons. In the clean limit
($\ell\to\infty$) $\lambda_0 = \lambda_{\rm L}$. The temperature
dependence of $\lambda$ within the BCS model is given implicitly by
the Eqs.(\ref{eq:K_bcs}),(\ref{eq:Lambda_BCS}), where again
$\lambda_0$ is used to indicate an effective London penetration
depth. The fits to the data are shown in Fig.\ref{fig:Pb_B_vs_z} and
were obtained in the following way: for the low-temperature
high-statistic data it was possible to fit $\lambda_0$ and $\xi_0$.
For the rest of the data, $\xi_0$ was fixed to the value found at
low temperature. This was necessary, since by approaching $T_c$, the
magnetic penetration profile becomes more and more exponential due
to the strong temperature dependence of $\lambda$. As a consequence,
the $\xi$-dependence of $B(z)$ weakens resulting in a drastically
growing uncertainty to determine the parameter.

\begin{figure}
  \centering
  \includegraphics[width=0.9\linewidth]{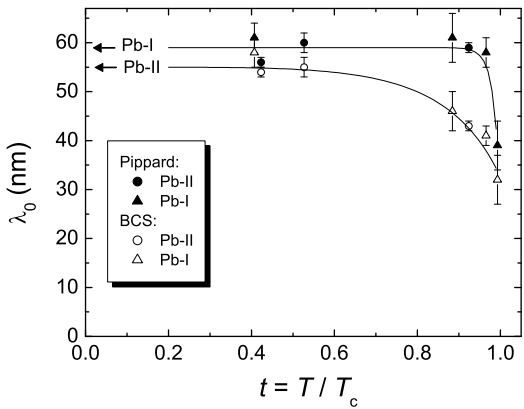}\\[-0.5cm]
  \caption{$\lambda_0$ determined at various temperatures $t=T/T_c$ for the Pb data.
           The temperature dependence of $\lambda$ in the Pippard model was chosen
           as $1/\lambda^2 \propto \sqrt{1-t^4}$, whereas in the BCS model
           the temperature dependence is given implicitly by
           Eqs.(\ref{eq:Bz}),(\ref{eq:K_bcs}),(\ref{eq:Lambda_BCS}). The curves
           are guides to the eyes.}\label{fig:Pb_lambda0_vs_t}
\end{figure}

In the models used to analyze $B(z)$ a temperature dependence of the
magnetic penetration $\lambda(T)$ is assumed (Pippard
Sec.\ref{sec:pippard}) or implicitly given (BCS Sec.\ref{sec:bcs}).
If these models describe the temperature dependence of $\lambda(T)$
correctly, $\lambda_0$ should be a \textit{constant} value within
the error bars for {\em all} the data sets.
Fig.\ref{fig:Pb_lambda0_vs_t} shows a graph where $\lambda_0$ is
plotted versus the reduced temperature $t=T/T_c$ for both models.
For the two-fluid temperature dependence, assumed in our Pippard
model, $\lambda_0$ is indeed temperature independent, except for
data close to $T_c$. The weak coupling BCS temperature dependence
for the magnetic penetration length, however, produces a clear
temperature dependent $\lambda_0$ meaning that the real temperature
dependence of the energy gap $\Delta(T)$ does not follow exactly the
weak coupling prediction. Table \ref{tab:results} shows the
collected results.

In the case of Nb (Fig.\ref{fig:Nb_B_vs_z}) the limited statistics of
the data do not allow us to fit $\xi_0$ directly. For the analysis it
was therefore fixed to its literature value (see Table
\ref{tab:results}). This system is at the borderline between
the non-local and local regimes. Therefore $B(z)$ was also analyzed with
a simple exponential model, leading to $\lambda_0^{\rm exp} =
33(3)$~(nm) which is similar to the non-local value given in Table
\ref{tab:results}. The $\chi^2$ suggests that also for Nb the
non-local regime is the appropriate one, but definitely higher
quality data are needed to resolve this issue.

\begin{figure}[t]
  \centering
  \includegraphics[width=\linewidth]{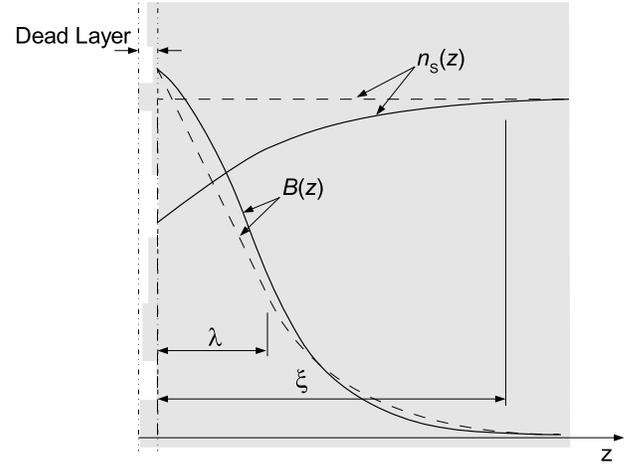}\\
  \caption{Effect on $B(z)$ of the suppression of $n_{\rm S}(z)$ at the surface.
           The shaded area represent the sample cross-section. The left side
           shows a surface with a terraces of typical size $\lesssim\xi_0$, which results
           in a ``dead layer''. The dashed lines show ideal superfluid density
           $n_{\rm S}(z)$ and magnetic penetration profile $B(z)$, respectively.
           The solid lines are the corresponding more realistic ones.}\label{fig:surface_roughness_dead_layer}
\end{figure}

The Ta data (Fig.\ref{fig:Ta_B_vs_z}) show a more pronounced
deviation from an exponential decay law. As for Nb it was necessary
in this case to fix $\xi_0$ to the literature value. Table
\ref{tab:results} summarizes the results.
In the data analysis we also considered the possibility of spurious
effects related to surface roughness or thickness variation of the
film due to the presence of terrace-like structures. If these
structures have a lateral size smaller or comparable to the
coherence length, the screening current will be very ineffective,
thus resulting in a dead layer at the surface. We effectively find a
better fit to the data assuming the presence of such a dead layer
(see Fig.\ref{fig:surface_roughness_dead_layer} and Table
\ref{tab:results}) in addition to the previously mentioned oxide
layer. The other case of surface terraces with lateral extension
$\gg$ $\xi_0$ would mimic a thickness distribution of the film.
Numerical simulations have shown that such a thickness variation
would only marginally affect the magnetic penetration depth.
A closer look at the Ta curves shows that compared to the other
systems the fit based on the BCS or Pippard model does reproduce the
data in a less satisfactory way. One possible origin for this
discrepancy is the neglect in both models of the suppression of the
supercurrent density on approaching the surface.
Fig.\ref{fig:surface_roughness_dead_layer} sketches the situation.
The models discussed in Sec.\ref{sec:theory} assume $n_{\rm S}$ to
be constant up to the surface (dashed line). However, from the
Ginzburg-Landau theory it is known (see e.g.
Ref.\onlinecite{poole95}) that close to the surface the order
parameter, and hence $n_{\rm S}$, are reduced. Such a reduction in
$n_{\rm S}$ makes the screening less effective, which, on top of the
non-local effects, would lead to a more pronounced deviation from
the exponential penetration profile than estimated by
Eq.(\ref{eq:Bz}) or its diffuse scattering counterpart. In the
presently investigated samples we expect such an effect to be more
pronounced in the cleaner superconducting Ta than in Pb (where also
small deviations from the theoretical curves are possibly present in
Fig.\ref{fig:Pb_B_vs_z}) because of its lower effective $\kappa$. In
order to quantify these effects in a comparison with our
experimental findings, it would be very useful to have a
self-consistent theoretical description of the physics outlined
above.

\begin{table}
  \centering
  \caption{Results of the analysis for Pb, Nb and Ta. The values for
           $\lambda_0^{\rm BCS, P}$ given, are the low temperature
           values. The ``dead layer'' corresponds to a shift of the $z$-origin of
           $B(z)$.}\label{tab:results}
  \begin{ruledtabular}
  \begin{tabular}{c|c|c|c|c}
    sample & compound   & $\lambda_0^{\rm BCS, P}$ & $\xi_0$                  & dead layer \\
           &            & (nm)                     & (nm)                     & (nm) \\ \hline\hline
    Pb-I   & Pb         & 59(3)                    & 90(5)                    & 6(1) \\
    Pb-II  & Pb         & 55(1)                    & 90(5)                    & 3(2) \\
    Nb     & Nb         & 27(3)                    & 39(fixed)\cite{poole95}  & 2(2)\\
    Ta     & Ta         & 52(2)                    & 92(fixed)\cite{poole95}  & 3(1)
  \end{tabular}
  \end{ruledtabular}
\end{table}

\section{Summary}\label{sec:summary}%

We have performed low energy muon spin rotation spectroscopy
experiments (\lem) on Pb, Nb and Ta films in the Meissner state. The
magnetic penetration profile $B(z)$ into the superconductor has been
determined on the nano-meter scale, thus providing a model
independent measure of the Meissner screening profile. $B(z)$ shows
clear deviations from the simple exponential decay law as expected
in the Pippard regime where the relation between the current density
and the vector potential is non-local in nature. Analyzing the data
within the Pippard and BCS model, the coherence length $\xi_0$ and
the London penetration length $\lambda_{\rm L}$ could be deduced.
Furthermore we could show that the two-fluid model approximation for
the temperature dependence of $\lambda \propto 1/\sqrt{1-(T/T_c)^4}$
is closer to the experiment than the BCS one which is $\lambda
\propto 1/\sqrt{K_{\rm BCS}(q\to 0, T, \ell\to\infty)}$. These
experiments based on the local profiling of the magnetic field
penetrating at the surface are the first measurements showing the
non-local nature of superconductivity in Pb, Nb and Ta on the
nanometer scale, and verify directly and quantitatively the
longstanding predictions of Pippard. $B(z)$ for Ta deviates
substantially from the simplest theoretical approach, which we
attribute to the neglect of the reduced superfluid density on
approaching the surface.

\section*{ACKNOWLEDGMENTS}%

This work was performed at the Swiss Muon Source, Paul Scherrer
Institute, Villigen, Switzerland. We thank M.~Doebeli for the RBS
measurements, D.~Eshchenko and D.~Ucko for the help during part of
the measurements. The long-term technical support by H.-P. Weber is
gratefully acknowledged.

\clearpage

%
% appendix
%----------------------------------------------------------------------------%
\appendix
\section{Modeling the determination of the magnetic field profile}\label{app:B_vs_z_reconstruction}%
\begin{figure}[t]
  \centering
  \includegraphics[width=0.9\linewidth]{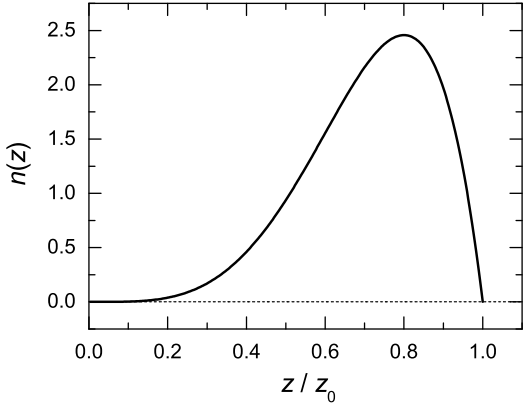}\\[-0.7cm]
  \caption{Model stopping distribution $n(z,E)$ of low energy
           muons.}\label{fig:nz}
\end{figure}

As pointed out in the paper the integral determination of $B(z)$
[Eq.(\ref{eq:nz_pB_int})] is the most effective method since it is
able to generate a complete curve from a single implantation energy.
As a cross-check of the use of the maximum entropy method to
determine $p(B,E)$ and $B(z)$ via Eq.(\ref{eq:nz_pB_int}), where the
determination of the statistical error is an unresolved
issue\footnote{Almost the same is true if Fourier transform would
have been used instead.}, we analyzed the $\mu$SR data directly in
the time domain which permits estimation of the statistical errors. The
time domain modeling used in our analysis has the disadvantage of
introducing systematic errors and only one of the discussed
approaches is acceptable for the investigation of non-local effects.

Two sets of models, which will be discussed separately in the
following paragraphs, can be treated analytically. The resulting
trend is similar in both cases, meaning that the ``mean value
determination'' (Sec.\ref{app:mean_value_reconstruction}) slightly
overestimates the real $B(z)$, whereas the ``peak value
determination'' (Sec.\ref{app:peak_value_reconstruction}) always
slightly underestimates $B(z)$. In the situation where one is
looking for non-local effects, characterized by an initial negative
curvature of $\log\left| B(z)/B_{\text{ext}}\right|$, the mean value
reconstruction is the method of choice because only it can produce
systematic errors with a positive curvature, and henceforth, in the
worst case, can only hide the presence of a negative curvature. All
these results are compiled in Fig.\ref{fig:B_vs_z_toy_model}.

\subsection{Mean Value Determination of $B(z)$}\label{app:mean_value_reconstruction}
The muon stopping distribution can only be calculated by Monte Carlo
codes. For the following discussion a simple stopping distribution
mimicking a realistic one is needed. We take as a model stopping
distribution $n(z,E)$

\begin{equation}\label{eq:nz_toy}
  n(z,E) = n_0 \, (z_0 - z) \, z^4,~~~ z \in [0,z_0],
\end{equation}

\noindent where $n_0 = 30\,z_0^{-6}$ is the normalization factor and
$z_0$ the maximum distance, which an implanted muon can reach. This
function is an acceptable approximation for a realistic implantation
profile (Fig.\ref{fig:nz}).

\noindent The mean value of $z$ is

\begin{equation}\label{eq:mean_z}
  \mean{z} = \frac{5}{7}\, z_0.
\end{equation}

\noindent Assuming further

\begin{equation}\label{eq:B_exp}
  B(z)=B_{\rm ext} \exp(-z/\lambda)
\end{equation}

\noindent one gets, utilizing the identity $n(z,E)\, \text{d}z =
p(B,E)\, \text{d}B$,

\begin{widetext}
\begin{eqnarray}\label{eq:mean_B_exp}
  \mean{B} &=& \int_0^{z_0} \text{d}z\, B(z) n(z,E) \nonumber \\
           &=& B_{\rm ext}\, n_0 \lambda^2 \left[ 24\lambda^3 (z_0 - 5 \lambda) +
               e^{-z_0/\lambda} \left\{ z_0^4 + 8 \lambda z_0^3 +
               36 \lambda^2 z_0^2 + 96 \lambda^3 z_0 + 120 \lambda^4
               \right\} \right] \\
           &=& B_{\rm ext} \exp(-\mean{z}/\lambda) +
           f(\mean{z}),~~~~~f(\mean{z}) \ge 0 \nonumber
\end{eqnarray}
\end{widetext}

\noindent with $\lim\limits_{\mean{z}\to 0} f(\mean{z}) = 0$ and
$f(\mean{z}) < \mean{B}(\mean{z})$ in the first two decades as shown
in Fig.\ref{fig:B_vs_z_toy_model}. Notice that the curvature of
$f(\mean{z})$ is positive.

In order to have an estimate not only for an exponential $B(z)$,
also the following $B(z)$ (resembling a non-local field profile
close to the solution in the extreme anomalous limit\cite{tinkham80}
$\xi\gg\lambda$) was analyzed

\begin{equation}\label{eq:extreme_anamalous_limit_B}
  B(z) = B_{\rm ext}\,\exp\left(-\frac{\sqrt{3}}{2}\, Q
  z\right)\,\cos\left(\frac{1}{2}\, Q z\right),
\end{equation}

\noindent where $Q = \left[\displaystyle\frac{\displaystyle 3 \pi}{\displaystyle
4}\, \displaystyle\frac{\displaystyle 1}{\displaystyle \lambda_{\rm L}^2 \xi_0}
\right]^{1/3}$. Also here an exact solution for the mean value can
be given

\begin{eqnarray}\label{eq:mean_B_ea}
  \mean{B}/B_{\rm ext} &=& c_0 + c_1 \exp\left(-\frac{7\sqrt{3}}{10}\, Q
       \mean{z}\right)\,\cos\left(\frac{7}{10}\, Q \mean{z}\right)
       + \nonumber \\
       && + c_2 \exp\left(-\frac{7\sqrt{3}}{10}\, Q \mean{z}\right)\,
       \sin\left(\frac{7}{10}\, Q \mean{z}\right)
\end{eqnarray}

\noindent where the coefficients $c_i$ are

\begin{widetext}
\begin{eqnarray*}
  c_0 &=& -\frac{1125000\, (7 \sqrt{3}\, Q \mean{z} - 50)}{117649\, (Q \mean{z})^6} \\
  c_1 &=& -\frac{375\,\left[150000 + 84000\, \sqrt{3}\, Q \mean{z} +
          44100\, (Q \mean{z})^2 - 2401\, (Q \mean{z})^4\right]}{117649\, (Q \mean{z})^6}\\
  c_2 &=& -\frac{375\, \left[ 12000 + 6300 \sqrt{3}\, Q \mean{z} +
            3920\, (Q \mean{z})^2 + 343 \sqrt{3} (Q \mean{z})^3\right]}{16807\,
            (Q \mean{z})^5}.
\end{eqnarray*}
\end{widetext}

\noindent These formulae look rather ugly but as can be seen in
Fig.\ref{fig:B_vs_z_toy_model}, the result is very similar to the
simpler case discussed before.

\subsection{Peak Value Determination of $B(z)$}\label{app:peak_value_reconstruction}
First lets assume the model stopping distribution from
Eq.(\ref{eq:nz_toy}). The peak value of the spatial coordinate is

\begin{equation}\label{eq:peak_z}
   z_p = \frac{4}{5}\, z_0.
\end{equation}

\begin{figure*}
  \centering
  \includegraphics[width=0.40\linewidth]{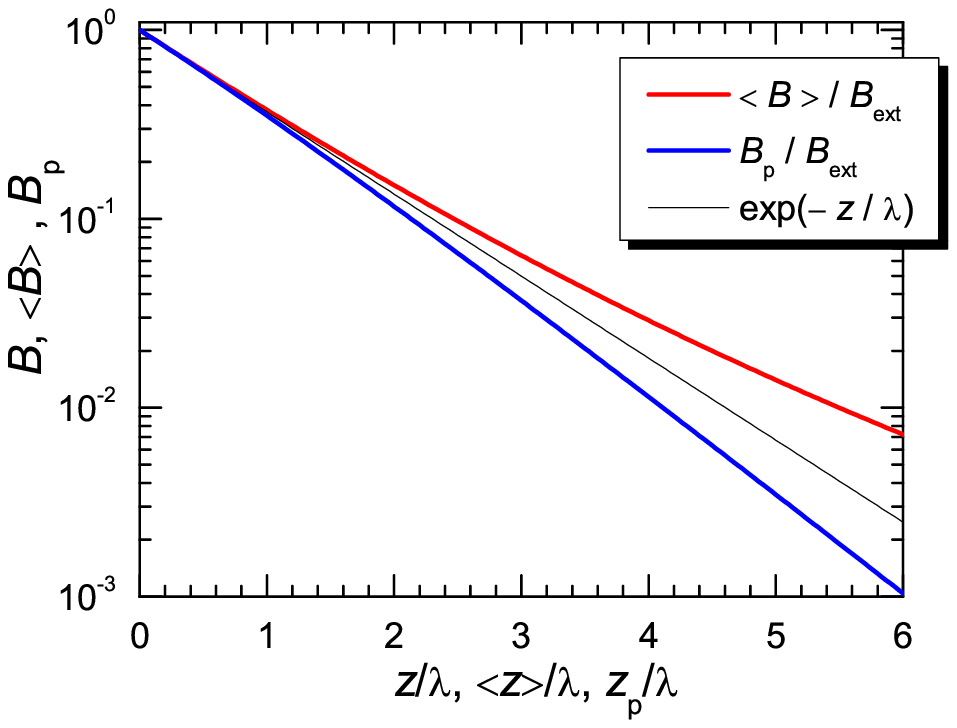}\qquad
  \includegraphics[width=0.45\linewidth]{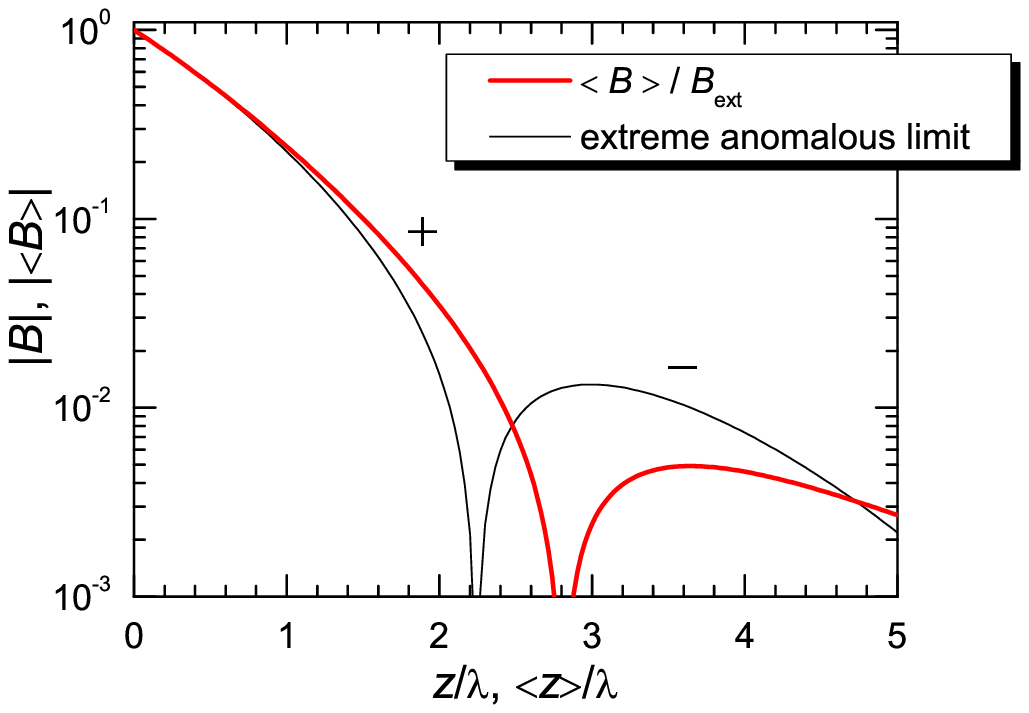}\\[-0.7cm]
  \caption{(Color online) Left: Analytic results for mean- and peak value
           determination assuming an exponential $B(z)$.\\
           Right: Extreme anomalous limit result for the mean value
           determination of $B(z)$.}\label{fig:B_vs_z_toy_model}
\end{figure*}

\noindent The corresponding peak position is obtained from

\begin{eqnarray*}
  p(B,E) &=& n(z,E) \, \left| \frac{dB}{dz} \right|^{-1} \\
         &=& \frac{\lambda}{B}\, (\lambda\,\ln(B_{\rm ext}/B))^4 \,
             \left( z_0 - \lambda\,\ln(B_{\rm ext}/B) \right),
\end{eqnarray*}

\noindent with $B \in [B_{\rm ext} e^{-z_0/\lambda}, B_{\rm ext}]$.
Since we are only interested in the peak position, the normalization
factor was suppressed. It follows that

\begin{eqnarray}\label{eq:peak_B_n_toy}
  B_p &=& B_{\rm ext} \exp\left[ \frac{-1}{2\lambda} \left(
          z_0 - 5\lambda + \sqrt{(z_0 + 5\lambda)^2 - 4 z_0}
          \right)\right] \nonumber \\
      &=& B_{\rm ext} \exp(-z_p / \lambda) - g(z_p),
\end{eqnarray}

\noindent with $g(z_p) \ge 0$ and $\lim_{z_p\to 0} g(z_p) = 0$.
The results found so far are compiled in
Fig.\ref{fig:B_vs_z_toy_model}.

Another ansatz, which is extremely close to the our analysis is the
following: Instead of assuming a model distribution $n(z,E)$
[Eq.(\ref{eq:nz_toy})], one assumes that $p(B,E)$ is Gaussian, i.e.\
the time dependent \lem signal is Gaussian damped.

\begin{equation}\label{eq:pB_Gauss}
  p(B,E) = \frac{1}{\sqrt{2\pi} \delta B}\, \exp\left[ - \frac{1}{2}\,
     \left(\frac{B-B_0}{\delta B}\right)^2 \right].
\end{equation}

\noindent with $B_0$ the position of the Gaussian peak. $\delta B$
is the width of the distribution. Assuming furthermore an
exponential $B(z)$ [Eq.(\ref{eq:B_exp})], one arrives at the
following stopping distribution

\begin{equation}\label{eq:nz_for_Gaussian_pB}
  n(z,E) = \frac{1}{\sqrt{2\pi} \lambda}\, \frac{B_0}{\delta B}
    \exp\left[ - \frac{1}{2}\, \left(\frac{B_{\rm ext}
    e^{-z/\lambda}-B_0}{\delta B}\right)^2 - \frac{z}{\lambda} \right]
\end{equation}

\noindent The peak values are than given by

\begin{eqnarray*}
  B_p &=& B_0 \\
  n_p &=& \lambda\, \ln\left[\frac{B_{\rm ext}}{2 \delta B^2}
          \left( \sqrt{B_0^2 + 4 \delta B} - B_0\right)\right],
\end{eqnarray*}

\noindent which leads to

\begin{equation}\label{eq:peak_B_p_toy}
  B_p(z_p) = B_{\rm ext}\, e^{-z_p/\lambda} \left[ 1 -
     \left(\frac{\delta B}{B_{\rm ext}}\right)^2
     e^{+2z_p/\lambda}\right],
\end{equation}

\noindent showing the same trend as in the previously discussed
case.

In conclusion, one finds that the peak value approach to determine
$B(z)$ leads to systematic errors that can mimic a magnetic
penetration profile of a non-local superconductor and therefore has
to be excluded from such an analysis. The mean value determination
of $B(z)$ also introduces systematic errors, however they tend to
diminish real non-local effects and are therefore less dangerous.

%
% bibliography
%------------------------------------------------------------------------%
%\bibliographystyle{prsty}
%\bibliography{../../bibliography/BiBListEM}

\begin{thebibliography}{10}

\bibitem{meissner33}
W. Meissner and R. Ochsenfeld, {Die Naturwissenschaften} {\bf 21},  787
  (1933).

\bibitem{london35}
F. London and H. London, {Proc. R. Soc. London, Ser. A} {\bf 149},  71  (1935).

\bibitem{pippard53}
A.~B. Pippard, Proc.~Roy.~Soc.~(London) {\bf A216},  547  (1953).

\bibitem{bcs57}
J. Bardeen, L.~N. Cooper, and J.~R. Schrieffer, Phys.~Rev. {\bf 108},  1175
  (1957).

\bibitem{suter04a}
A. Suter, E. Morenzoni, R. Khasanov, H. Luetkens, T. Prokscha, and N.
  Garifianov, Phys.~Rev.~Lett. {\bf 92},  087001  (2004).

\bibitem{sommerhalder61a}
R. Sommerhalder and H. Thomas, Helv.~Phys.~Acta {\bf 34},  29  (1961).

\bibitem{sommerhalder61b}
R. Sommerhalder and H. Thomas, Helv.~Phys.~Acta {\bf 34},  265  (1961).

\bibitem{drangeid62}
K.~E. Drangeid and R. Sommerhalder, Phys.~Rev.~Lett. {\bf 8},  467  (1962).

\bibitem{doezema84}
R.~E. Doezema, J.~N. Huffaker, S. Whitmore, J. Slinkman, and W.~E. Lawrence,
  Phys.~Rev.~Lett. {\bf 53},  714  (1984).

\bibitem{nutley94}
M.~P. Nutley, A.~T. Boothroyd, C.~R. Staddon, D.~M. Paul, and J. Penfold,
  Phys.~Rev.~B {\bf 49},  15789  (1994).

\bibitem{schrieffer64}
J.~R. Schrieffer, {\em Theory of Superconductivity} (Addison Wesley, San
  Francisco, 1964).

\bibitem{tinkham80}
M. Tinkham, {\em Introduction to Superconductivity} (Krieger, New York, 1980).

\bibitem{kosztin97}
I. Kosztin and A.~J. Leggett, Phys.~Rev.~Lett. {\bf 79},  135  (1997).

\bibitem{muehlschlegel59}
{B.~M{\"u}hlschlegel}, Z.~Phys. {\bf 155},  313  (1959).

\bibitem{halbritter71}
J. Halbritter, Z.~Phys. {\bf 243},  201  (1971).

\bibitem{parks69}
{\em Superconductivity}, edited by R. Parks (Dekker, New York, 1969).

\bibitem{carbotte90}
J.~P. Carbotte, Rev.~Mod.~Phys. {\bf 62},  1027  (1990).

\bibitem{nam67}
S.~B. Nam, Phys.~Rev. {\bf 156},  470  (1967).

\bibitem{halbritter87}
J. Halbritter, Appl.~Phys.~A {\bf 43},  1  (1987).

\bibitem{schenck85}
A. Schenck, {\em Muon Spin Rotation Spectroscopy: Principles and Applications
  in Solid StatePhysics} (Adam Hilger, Bristol, 1985).

\bibitem{morenzoni94}
E. Morenzoni, F. Kottmann, D. Maden, B. Matthias, M. Meyberg, T. Prokscha, T.
  Wutzke, and U. Zimmermann, Phys. Rev. Lett. {\bf 72},  2793  (1994).

\bibitem{morenzoni00}
E. Morenzoni, H. Gl{\"u}ckler, T. Prokscha, H.~P. Weber, E.~M. Forgan, T.~J.
  Jackson, H. Luetkens, C. Niedermayer, M. Pleines, M. Birke, A. Hofer, F.~J.
  Litterst, T. Riseman, and G. Schatz, Phys.\ B {\bf 289-290},  653  (2000).

\bibitem{wu97}
N. Wu,  in {\em The Maximum Entropy Method}, Vol.~32 of {\em Springer Series in
  Information Science}, edited by T.~S. Huang, T. Kohonen, and M.~R. Schroeder
  (Springer, Berlin, 1997).

\bibitem{skilling84}
J. Skilling and R.~K. Bryan, Mon.\ Not.\ R.\ astr.\ Soc. {\bf 211},  111
  (1984).

\bibitem{rainford94}
B.~D. Rainford and G.~J. Daniell, Hyperfine Interact. {\bf 87},  1129  (1994).

\bibitem{riseman00b}
T. Riseman and E.~M. Forgan, Physica~B {\bf 289-290},  718  (2000).

\bibitem{eckstein91}
W. Eckstein, {\em Computer Simulation of Ion-Solid Interactions} (Springer
  Verlag, Berlin, 1991).

\bibitem{morenzoni02}
E. Morenzoni, H. Gl{\"u}ckler, T. Prokscha, R. Khasanov, H. Luetkens, M. Birke,
  E.~M. Forgan, and C. Niedermayer, Nucl. Instr. Meth. B {\bf 192},  254
  (2002).

\bibitem{jackson00b}
T.~J. Jackson, T.~M. Riseman, E.~M. Forgan, {H.~Gl{\"u}ckler}, T. Prokscha, E.
  Morenzoni, M. Pleines, C. Nidermayer, G. Schatz, H. Luetkens, and J.
  Litterst, Phys.~Rev.~Lett. {\bf 84},  4958  (2000).

\bibitem{yip92}
S.~K. Yip and J.~A. Sauls, Phys.~Rev.~Lett. {\bf 69},  2264  (1992).

\bibitem{amin98}
M.~H.~S. Amin, I. Affleck, and M. Franz, Phys.~Rev.~B {\bf 58},  5848  (1998).

\bibitem{poole95}
C.~P. Poole, H.~A. Farach, and R.~J. Creswick, {\em Superconductivity}
  (Academic Press, San Diego, 1995).

\end{thebibliography}

\end{document}